\documentclass[12pt]{article}
\normalbaselines
\catcode `\@=11
\@addtoreset{equation}{section}

\def\section{\@startsection {section}{1}{\z@}{-3.5ex plus -1ex minus
     -.2ex}{2.3ex plus .2ex}{\normalsize\bf}}
\def\subsection{\@startsection{subsection}{2}{\z@}{-3.25ex plus -1ex minus
 -.2ex}{1.5ex plus .2ex}{\normalsize\bf}}

\def\thebibliography#1{\section*{References\markboth
  {REFERENCES}{REFERENCES}}\list
  {[\arabic{enumi}]}{\settowidth\labelwidth{[#1]}\leftmargin\labelwidth
  \advance\leftmargin\labelsep
  \usecounter{enumi}}
  \def\newblock{\hskip .11em plus .33em minus -.07em}
  \sloppy
  \sfcode`\.=1000\relax}
 
\catcode `\@=12
\newcommand{\be}{\begin{equation}}
\newcommand{\ee}{\end{equation}}
\newcommand{\bea}{\begin{eqnarray}}
\newcommand{\eea}{\end{eqnarray}}

\begin{document}

\vspace*{2.5cm}
\noindent
{ \bf EVOLUTION LOOPS AND SPIN-$1/2$ SYSTEMS}\vspace{1.3cm}\\
\noindent
\hspace*{1in}
\begin{minipage}{13cm}
David J. Fern\'andez C.${}^{1}\dagger$ and Oscar Rosas-Ortiz${}^{1,2}\ddagger
$ \vspace{0.5cm}\\
 $^{1}$ Departamento de F\'{\i}sica, CINVESTAV-IPN \\
\makebox[3mm]{ }A.P. 14-740, 07000 M\'exico D.F., MEXICO \\
\vspace{0.1cm} \\
 $^{2}$ Departamento de F\'{\i}sica Te\'orica\\
\makebox[3mm]{ }Universidad de Valladolid\\ 
\makebox[3mm]{ }47011 Valladolid, SPAIN\\
\vspace{0.1cm} \\
\makebox[3mm]{ }$\dagger$E-mail:  david@fis.cinvestav.mx\\
\vspace{0.1cm} \\
\makebox[3mm]{ }$\ddagger$E-mail:  orosas@klander.fam.cie.uva.es\\
\makebox[3mm]{ }\hskip1.7cm orosas@fis.cinvestav.mx
\end{minipage}
%
%

\vspace*{0.5cm}

\begin{abstract}
\noindent
The derivation of a new family of magnetic fields inducing exactly
solvable spin evolutions is presented. The conditions for which these
fields generate the {\it evolution loops\/} (dynamical processes for which
any spin state evolves cyclically) are studied. Their natural connection
with geometric phases and the corresponding calculation is also
elaborated. 
\end{abstract}

\section{\hspace{-4mm}.\hspace{2mm}INTRODUCTION}

In nonrelativistic quantum mechanics, the pure states of spin-$1/2$
systems are usually represented by two-component ket vectors $\vert \psi
\rangle \equiv (\psi_1, \psi_2)$, where $\psi_1, \psi_2 \in {\bf C}^1$ are
the components of $\vert\psi\rangle$ along the two orthogonal eigenstates
$\vert +\rangle$ and $\vert - \rangle$ of $S_z$ with eigenvalues $\hbar/2$
and $-\hbar/2$ respectively. If the spin is placed in a homogeneous
time-dependent magnetic field ${\bf B}(t)$, then the Hamiltonian is
described by $H(t) \equiv -\mu{\bf B}(t)\cdot{\bf S}$, and the
Schr\"{o}dinger equation governing the evolution becomes explicitly
time-dependent: 
\be
i\hbar \frac{d}{dt} \vert \psi(t) \rangle= H(t)\, \vert \psi(t) \rangle =
-\mu{\bf B}(t)\cdot{\bf S} \vert \psi(t) 
\rangle . 
\label{I1}
\ee

In order to solve (\ref{I1}), usually it is rewritten in terms of the
time-evolution operator $U(t)$, $U(t=0) \equiv I$. The most elementary
case arises when $[H(t), H(t')]=0$ for all $t \neq t'$ \cite{Sak85}. In
the case when $[H(t), H(t')] \neq 0$, however, to sum up the {\it
continuous\/} Baker-Campbell-Hausdorff exponent becomes hard \cite{Mie70},
and thus it is not easy to construct exact solutions to (\ref{I1}). Due to
this, the approximate methods (see for example \cite{Coh77} and references
quoted therein) or direct numerical techniques \cite{Shi65} are the
standard tools applied in the general case. Is there any optional
technique leading to exactly solvable situations? The answer seems to come
of the so called {\it inverse techniques}, which take full advantage of
the geometrical picture involved in the spin-$1/2$ description
\cite{Fer97}.

Suppose that somehow one knows the time evolution of the system, {\it
i.e.}, the state vector $\vert\psi(t)\rangle$ satisfying (1.1) for all
$t$, but there is no information about the external field ${\bf B}(t)$
driving the system. Can one find at least one field which could in
principle be created and would dynamically induce that state? This is the
essence of the {\it dynamical manipulation problem\/} whose main ideas
were developed by Lamb \cite{Lam69} and followed later by other authors
\cite{Lub74,Mie77,Fer92}. In the spin-1/2 case (1.1), this technique was
successfully applied to generate exactly solvable situations \cite{Fer97}. 
It would be interesting to show that this is so using the standard direct
approach. 

The aim of this paper is to prove explicitly that given the family of
magnetic fields derived in \cite{Fer97}, they lead to exact solutions to
equation (1.1). Moreover, we shall show that those magnetic fields can
induce some special dynamical processes on the system such that the
evolution operator becomes the identity (modulo phase)  at some $t=\tau$,
{\it i.e.}, $U(\tau) = e^{i\phi}I$. Such kind of processes have been
widely studied under the name of {\it evolution loops\/} (EL). It has been
also proposed that the EL can be {\it perturbed} in order to induce any
unitary operator as a result of the precession of the distorted loop
\cite{Mie77}. This proposal has been successfully realized by inducing the
squeezing inside of a modified Penning trap and by rigidly displacing the
wavepacket in a {\it magnetic chamber} perturbed by homogeneous
time-dependent electric fields \cite{Fer92}. Here, the EL will mean just
closed orbits on $S^2$ (independent of the initial condition) which can be
either periodic or aperiodic and whose corresponding geometric phases can
be simply evaluated.

The paper is organized as follows. In Section 2 we briefly sketch the
derivation of the magnetic fields of \cite{Fer97} with a discussion of
their basic properties. In section 3 we directly solve (\ref{I1}) taking
for ${\bf B}(t)$ the analytic expressions of these fields. Then we shall
establish the conditions on the fields in order to induce the EL on the
system.  To conclude, we will show that these EL give place to integrable
expressions for the corresponding geometric phases.

\section{\hspace{-4mm}.\hspace{2mm}THE INVERSE METHOD}

For the set of pure states of spin-$1/2$ systems, a geometric picture is
easily found by noticing that the inner product between any two ket
vectors is invariant if both are multiplied by an arbitrary unimodular
complex number $\lambda \in {\bf C}^1$, $\vert \lambda \vert =1$. Hence,
the space of physical states corresponds to the projective space ${\bf
C}P(2)$ which, as is well known, is usually modeled by the Riemman sphere
$S^2$ with each point of the surface representing a pure state of the
spin-$1/2$ system.

Now, according to Eherenfest theorem the mean value of the operator ${\bf
S} \equiv S_x \, {\bf i} + S_y \, {\bf j} + \, S_z {\bf k}$, where $S_k
\equiv (\hbar/2) \, \sigma_k$ and $[ S_k , S_{\ell}] = i \hbar \,
\epsilon_{k \ell m} S_m$, evolves as: 
\be
\frac{d}{dt} \langle {\bf S} \rangle = - \mu {\bf B}(t) \times \langle
{\bf S} \rangle = - {\bf b}(t) \times \langle {\bf S} \rangle, 
\label{2.1}
\ee
where ${\bf b}(t) \equiv \mu {\bf B}(t)$. The identification ${\bf n}
\equiv (2/\hbar)\langle {\bf S} \rangle$, ${\bf n} \cdot {\bf n} = 1$
shows that (\ref{2.1}) is precisely the dynamical rule governing the
evolution on $S^2$. In the direct approach the initial vector ${\bf n}(0)$
and the field ${\bf b}(t)$ are given, and one looks for the solution ${\bf
n}(t)$ to (2.1). Here we assume that the spin state ${\bf n}(t)$ is given
and rewrite (2.1) as $\dot{\bf n}(t)= M[ {\bf n}(t) ] {\bf b}(t)$. An
element of arbitrariness arises by noticing that the matrix $M[ {\bf n}(t) 
]$ is antisymmetric, and hence its determinant is equal to zero. Thus
$M^{-1}[ {\bf n}(t)]$ does not exist, which does not allow to determine
uniquely ${\bf b}(t)$. Let us take the third component $b_3(t)$ also as
given; henceforth the other two components become: 
\be
b_1(t) = [b_3(t)n_1(t) + \dot{n}_2(t)]/ n_3(t) , \qquad
b_2(t) = [b_3(t)n_2(t) - \dot{n}_1(t)]/ n_3(t).
\label{1.2}
\ee

Notice that, departing from a given point ${\bf n}(0)\in S^2$, any other
point ${\bf n}(t) \in S^2$ can be achieved by a set of successive
infinitesimal rotations encoded in $R(t) \in SO(3)$, {\it i.e.}, ${\bf
n}(t)=R(t){\bf n}(0)$. Therefore, the field ${\bf b}(t)$ given by (2.2)
depends on the generic motion (a generalized rotation) and the initial
condition.  Thus, two paths with common $R(t)$ but different ${\bf n}(0)$
are induced by two different fields with the same $b_3(t)$. Would it be
possible that trajectories sharing the same $R(t)$ determine a unique
${\bf b}(t)$? In order to find the answer, let us consider the case when
${\bf n}(t)$ rotates simultaneously around two fixed directions with
variable angular velocities. By simplicity, one of these directions is
fixed along ${\bf k}$ and the other one along a vector ${\bf e}_\chi$ on
the $x-z$ plane at an angle $\chi$ from ${\bf k}$, ${\bf e}_\chi =
\sin\chi{\bf i} + \cos\chi{\bf k}$. The rotation matrix is:
\be
\label{rot}
R(t) = R_3(\beta(t))R_\chi(\alpha(t)) = R_3(\beta(t)) R_2^{-1}(-\chi) 
R_3(-\alpha(t))R_2(-\chi) ,
\ee
where $R_2(\omega)$ and $R_3(\omega)$ are finite rotations by $\omega$
around ${\bf j}$ and ${\bf k}$ respectively and $\alpha (0) = \beta(0) =
0$. Using (\ref{1.2}), one will find a ${\bf b}(t)$ dependent of the
initial condition; the field independent of ${\bf n}(0)$ arises after
imposing the restriction $b_3(t) + \mathaccent95 \beta(t) = \mathaccent95
\alpha(t)\cos\chi,$ and then: 
\be
{\bf b}(t) = \dot{\alpha}(t) \sin\chi \,\bigl[\cos \beta(t){\bf i} + \sin
\beta(t){\bf j}\bigr] + \bigl[\dot{\alpha}(t)\cos\chi -
\dot{\beta}(t)\bigr]{\bf k}. 
\label{1.3}
\ee

Let us remark that, due to its dependence on two arbitrary functions,
equation (\ref{1.3}) represents a wide family of analytically solvable
fields inducing the rotation described by (2.3) on any initial vector
${\bf n}(0)$.  An interesting result comes out by analyzing the following
physical situation. Suppose that the spin initially points along an
arbitrary direction ${\bf n}(0) = {\bf e}_+$; what is the probability that
at time $t$ the spin will be in the corresponding orthogonal state ${\bf
e}_-$?  Notice that orthogonal vectors $\vert \psi \rangle$ on the Hilbert
space ${\cal H}$ correspond to {\it antipodal} points on the sphere $S^2$.
By simplicity, let us choose ${\bf e}_{\pm} \equiv (0,0, \pm 1)$; then the
probability transition is given by:
\be
P_{+\rightarrow -}(t) = [1-n_3(t)]/2 = \sin^2\chi\sin^2[\alpha(t)/2].   
\label{1.4}
\ee

As our treatment is exact, equation (\ref{1.4}) is indeed a generalization
of Rabi's formula for any $t$ and arbitrary $\alpha(t)$. It reduces to the
standard Rabi expression when we take $\alpha(t) = \alpha_0 t$ for small
$t$ (see {\it e.g.\/}, Rabi {\it et.al.\/} \cite{Rab54} and Shirley
\cite{Shi65}).  Therefore, by choosing specific forms for $\alpha(t)$ and
$\beta(t)$ in (\ref{1.3}) one is led to obtain diverse particular cases of
${\bf b}(t)$ \cite{Fer97}, some of which could have been previously
discussed. 

Let us finish this section by remarking that the product $b_k(t) S_k$ of
(\ref{I1}) satisfy $b_k(t)b_{\ell}(t') \, [S_k,S_{\ell}] = i \hbar \,
\epsilon_{k \ell m} b_k(t)b_{\ell}(t') S_m$. Hence, for arbitrary
functions $\alpha(t)$ and $\beta(t)$ (see the expressions of $b_k(t)$ in
(\ref{1.3})), the Hamiltonians $H(t)= -{\bf b}(t) \cdot {\bf S}$ at
different times do not commute, {\it i.e.}, $[H(t), H(t')] \neq 0$, $t
\neq t'$. Then, we have arrived at a family of exactly solvable
Hamiltonians which, at first sight, should be solved by approximate or
numerical methods when using the direct approach.

\section{\hspace{-4mm}.\hspace{2mm}THE DIRECT METHOD}

In this section we are going to solve (\ref{I1}) with the magnetic field
(\ref{1.3}) using ordinary quantum mechanical operator methods. The key
point is that, by means of the commutation rules $[\sigma_k, \sigma_\ell]
= 2i\epsilon_{k\ell m}\sigma_m$, and the expression
\[
e^{-i \beta(t) \, \sigma_3 /2} \, \sigma_1 \, e^{i \beta (t) \, \sigma_3 /
2} = \cos \beta(t) \, \sigma_1 + \sin \beta(t) \, \sigma_2,
\]
the Hamiltonian $H(t)=-{\bf b}(t) \cdot {\bf S}$ can be rewritten as:
\be
\label{3.1}
H(t) = e^{- i \beta(t) \, \sigma_3 / 2} \, [ H_{eff}(t) + \hbar
\dot\beta(t) 
\sigma_3 /2 ] \, e^{i \beta (t) \, \sigma_3 / 2},
\ee
where $H_{eff}(t)$ is defined by
\be
\label{3.2}
- (2/\hbar) \, H_{eff}(t) \equiv \dot\alpha(t) ( \sin \chi \sigma_1 + \cos
\chi \sigma_3 ). 
\ee
Now, let us introduce a new reference frame which rotates with angular
velocity $\dot\beta(t)$ around ${\bf k}$. Notice that rotating frames are
very useful in magnetic resonance because when successfully used, the
original problem can be mapped into a static one, which simplifies the
physical analysis of the problem \cite{Rab54}. The key transformation in
this case is given by $U(t) = e^{-i\beta(t) \, \sigma_3 /2}\, W(t)$, where
$W(t)$ is a new unitary operator satisfying the equation: 
\be
\label{3.3}
i \hbar \, \frac{d W(t)}{dt} =  H_{eff}(t) W(t).
\ee
It is clear now that $H_{eff}(t)$ is the Hamiltonian in the rotating
frame, and thus $W(t)$ represents precisely the evolution operator in that
frame. Notice that $H_{eff}(t)$ satisfies $[ H_{eff}(t), H_{eff}(t')] =
0$; henceforth, the solution of (\ref{3.3}) is given by just integrating
$H_{eff}(t)$: 
\bea
\label{3.4}
W(t) &=& e^{(-i/\hbar) \int_{0}^t H_{eff}(t') \, dt'} = e^{i\alpha(t) (
\sin \chi \sigma_1 + \cos \chi \sigma_3 )/2}\\
\nonumber 
&=& \cos (\alpha(t)/2)  + i (\sin \chi \sigma_1 + \cos \chi \sigma_3 )
\sin (\alpha(t)/2),
\eea
where we have used again the algebraic properties of $\sigma_k$. Notice
that at $t=0$ equation (\ref{3.4}) reads $W(0) = I$. Hence, initially the
evolution operators $U(t)$ and $W(t)$ coincide. At arbitrary times $t \neq
0$, there will be a time-dependent factor operator
$\exp(-i\beta(t)\sigma_3/2)$ making the difference between the
descriptions at the lab and at the rotating frames.

Let us construct now the generic expression for the ket vectors $\vert
\psi(0) \rangle$. First, the normalization condition gives $\vert
\psi_1(0)  \vert^2 + \vert \psi_2(0)  \vert^2 = 1$. It suggest the
following representation $\psi_1(0)= \cos(\theta_0/2)$, $\psi_2(0)=
\sin(\theta_0/2)$.  Now, the normalization remains invariant if we take
$\psi_1(0) = \exp (-i \phi_1)\cos(\theta_0/2)$, $\psi_2(0) = \exp (i
\phi_2)\sin(\theta_0/2)$. As mentioned at Section 2, the multiplication of
$\vert\psi(0)\rangle$ by a common phase factor $\lambda =
e^{i(\phi_1-\phi_2)/2}$ does not change the representative of the
corresponding physical state.  Hence, we can take: 
\be
\label{3.5}
\vert \psi(0) \rangle \equiv \left(
\begin{array}{c}
\psi_1(0)\\
 \\
\psi_2(0)
\end{array}
\right) = \left(
\begin{array}{l}
\cos (\frac{\theta_0}{2}) \, e^{-i \phi_0/2}\\
 \\
\sin (\frac{\theta_0}{2}) \, e^{i \phi_0/2}
\end{array}
\right),
\ee
with $\phi_0 = \phi_1 + \phi_2$. We are using the half angle convention
whose utility will be apparent below. The solutions $\vert \psi (t)
\rangle$ to (\ref{I1}), with the magnetic field ${\bf b}(t)$ given in
(\ref{1.3}), arise just as time displacements induced by the evolution
operator $U(t)$ acting on the initial vector (\ref{3.5}):
\be
\label{3.6}
\vert \psi (t) \rangle = U(t) \, \vert \psi (0)  \rangle = e^{-i\beta(t) 
\, \sigma_3 /2} \, e^{i\alpha(t) ( \sin \chi \sigma_1 + \cos \chi \sigma_3
)/2} \vert \psi (0)  \rangle,
\ee
where we have used (\ref{3.4}). Let us remark that the map $\vert \psi(t) 
\rangle \rightarrow (2/\hbar) \langle {\bf S} \rangle (t) = {\bf n}(t)$
reproduces the results derived in Section 2, {\it i.e.}, the vector ${\bf
n}(t)$, connected with $\vert \psi(t) \rangle$ in (\ref{3.6}) by this map,
is the result of two simultaneous rotations performed by the vector
\be
\label{initial}
{\bf n}(0) \equiv (2/\hbar)\, \langle \psi(0) \vert \, {\bf S} \, \vert
\psi(0) \rangle = (\sin \theta_0 \cos \phi_0, \, \sin \theta_0 \sin
\phi_0, \, \cos \theta_0) 
\ee
around the two directions and with the two angular velocities
characteristic of the matrix (\ref{rot}). Notice that, from a geometrical
point of view, the above map corresponds to the {\it Hopf map}
\cite{Hop31,Nak90}, which is useful in the description of monopole
magnetic charges in the fibre-bundle formulation of electrodynamics
\cite{Ryd80}, and provides an interesting geometrical interpretation of
the Aharonov-Bohm effect \cite{Aha59,Nak90}. 


\subsection{\hspace{-5mm}.\hspace{2mm}Evolution Loops and Geometric
Phases}

Let us formulate now the requirements which has to be satisfied in order
to induce the evolution loops. With this aim, let us rewrite the evolution
operator in (\ref{3.6}) as: 
\[
U(t) = \left[ \cos \left( \frac{\beta(t)}{2} \right) - i \sigma_3 \sin
\left( \frac{\beta(t)}{2} \right) \right] \left[ \cos \left(
\frac{\alpha(t)}{2} \right) + i ( \sin \chi \sigma_1 + \cos \chi \sigma_3
) \sin \left( \frac{\alpha(t)}{2} \right) \right].
\]
It is apparent that $\alpha(\tau) = 2 \ell \pi$ and $\beta(\tau) = 2 m
\pi$, with $\ell, m \in {\bf Z}$, produce $U(\tau) = \cos (m \pi) \cos
(\ell \pi) = \cos (m + \ell) \pi = (-1)^{m + \ell}I$. The {\it loop
conditions} are thus: 
\be
\label{3.7'}
\alpha(\tau) = 2 \ell \pi,\quad \beta(\tau) = 2 m \pi, \quad 
m,\ell \in {\bf Z}; \, \tau >0.
\ee
Notice that there is a {\it strong} loop condition in which the evolution
operator becomes the identity {\it sensu stricto}. In our spin-1/2 case
this strong loop condition consists of the restriction (3.8) and the
additional requirement $m + \ell=2k, \ k\in{\bf Z}$. Here and throughout
the paper we will use the {\it relaxed} loop condition (3.8). 

The conditions (\ref{3.7'}) have been intuitively used in the case when
the magnetic field (2.4) rotates with constant angular velocity, and then
the vector ${\bf n}(t)$ describes a {\it hypocycloid} on $S^2$
\cite{Fer992} (similar results can be found in \cite{Skr94} and Zhang {\it
et. al.\/} \cite{Shi65}). That case is recovered here by taking $\alpha(t) 
= \alpha_0 t$, $\beta(t) = \beta_0 t$ and by forcing the loop condition
(\ref{3.7'}). Other selection of the functions $\alpha(t)$ and $\beta(t)$
allows us to generate deformed versions of such a case and generalized
nontrivial cases (see the specific examples of \cite{Fer97}).

The loop condition (\ref{3.7'}) and equation (\ref{3.6}) give $\vert \psi
(\tau) \rangle = (-1)^{m+\ell}\vert \psi (0)\rangle$. As Aharonov and
Anandan have shown, any cyclic quantum state has naturally associated a
geometric phase $\gamma$ characterizing somehow the projective Hilbert
space curvature (see also the general geometric treatment introduced long
ago by Mielnik \cite{Mie68}). In the spin-1/2 case, it turns out that
$\gamma = -\Delta\Omega/2$, where $\Delta\Omega$ is the solid angle
subtended by the oriented closed curve ${\bf n}(t)$:
\be
\label{3.9}
\Delta\Omega = \int_0^\tau\frac{n_1{\dot n}_2 -n_2{\dot n}_1}{1+n_3}dt. 
\ee

Let us remark that, although the general expression (\ref{3.9}) has been
recurrently studied and discussed in the literature [15-20], it is not
always possible to perform the involved integrals (see \cite{Ana97} and
references quoted therein). Therefore, it is interesting to look for
explicit expressions for $\gamma$. For the magnetic fields (2.4) we have
gotten the generic time-evolution (\ref{3.6}), and this added to the loop
conditions (\ref{3.7'}) allow one to simplify considerably the
calculation:
\be
\label{3.10}
\gamma = [\ell - m + \cos(\theta_0 - \chi) ( m \cos\chi - \ell)] \pi
-\frac{1}{2} \sin \chi \sin (\theta_0 - \chi) \int_0^\tau{ \dot\beta} (t) 
\cos \alpha(t) dt,
\ee
where in (\ref{initial}) we have taken $\phi_0 = 0$.  Notice the arising
of the atypical integral term in (\ref{3.10}). This formula generalizes
the corresponding expression for the traditional rotating magnetic field
with constant amplitude and angular velocity \cite{Fer992}.

In order to illustrate the generality of our expression (\ref{3.10}), let
us consider some particular cases of (\ref{1.3}). The first simple case
arises by taking $b_3(t) = b_0$ in (\ref{1.3}), and thus
\be 
\label{3.11} 
\gamma = [\ell -m + \cos(\theta_0 - \chi) ( m \cos\chi - \ell)] \pi
+\frac{b_0}{2} \sin \chi \sin (\theta_0 - \chi) \int_0^\tau \cos \alpha(t) 
dt.
\ee 
The remaining integral depends just on $\alpha(t)$ and $\tau$. Notice that
it vanishes for the simplest time-dependent function $\alpha(t)  =
\alpha_0 t$ with the loop condition $\alpha_0\tau = 2 \ell\pi$, which is
indeed the case discussed in \cite{Fer992} and \cite{Skr94}. The first
nontrivial case of (\ref{3.11}) arises after taking $\alpha(t)$ quadratic
in $t$, for instance, $\alpha (t) = \alpha_0 t^2$, with $\alpha_0 = 5/2
\pi$, $b_0=3$ and $\cos \chi = 4/5$. This choice immediately satisfies the
loop condition $\alpha (\tau = 2 \pi)  = 10 \pi$ and $\beta (2 \pi) = 2
\pi$, and leads to $\int_0^{2 \pi} \cos \alpha (t)  dt = \pi C (2
\sqrt{5})/ \sqrt{5} = 0.700896 \neq 0$, where $C(x)$ is the Fresnel cosine
integral.

Up to now, we have seen that the loop conditions (\ref{3.7'}) guarantee
the cyclic time evolution of any initial state (\ref{3.5}) (or
equivalently (\ref{initial})). It should be clear now that without these
restrictions on $\alpha(t)$ and $\beta(t)$, an arbitrary state not
necessarily will be cyclic. An interesting question arises: are the loop
conditions (\ref{3.7'}) the only way to ensure cyclic evolutions of the
involved states? In order to get an answer let us consider two special
initial states of the spin-$1/2$ system. Let us make in (3.7) $\phi_0 = 0$
and $\theta_0 = \chi$, and denote the resulting vector by ${\bf
n}_{\chi}^{+} (0)$; now let us make $\phi_0 = \pi$ and $\theta_0 =
\pi-\chi$, and denote the resulting vector by ${\bf n}_{\chi}^{-} (0)$.
The corresponding evolution (\ref{3.6}) leads to ${\bf n}_{\chi}^{\pm}
(t)$, where ${\bf n}_{\chi}^{\pm} (0) \equiv \pm {\bf e}_{\chi}$. Hence,
if the spin state points initially along $\pm {\bf e}_{\chi}$, the
rotataion around that vector has no effect on it, and we have ${\bf
n}_\chi^+(t) = \sin\chi\cos\beta(t){\bf i} + \sin\chi\sin\beta(t){\bf j} +
\cos\chi{\bf k}$, ${\bf n}_\chi^-(t) = -{\bf n}_\chi^+(t)$. At times
$T_n$, such that $\beta(T_n) = 2n\pi$, these states turn back to ${\bf
n}_\chi^\pm(0)$. The subtended solid angles are $\Delta\Omega^\pm =
2n\pi(1 \mp \cos\chi)$, and so the geometric phases become: 
\be
\label{3.12}
\gamma^\pm = -n\pi(1\mp\cos\chi).
\ee
Let us notice that $n$ represents here the number of effective turns that
${\bf n}_\chi^\pm(t)$ performs around the $z$-axis, and that (\ref{3.12}) 
is a general result valid for any $\chi$, with $\alpha(t)$ and $\beta(t)$
arbitrary.  In particular, when $\chi = \pi/2$ (orthogonal rotations) we
get $\gamma^\pm= -n\pi$;  this phase is zero (modulo $2\pi$) when $n$ is
even, while it is $\pi$ (modulo $2\pi$) when $n$ is odd. On the other
hand, if the number of effective turns is zero ($\chi$ arbitrary) we get
$\gamma^\pm = 0$. This is indeed the trivial case discussed by Zhang {\it
et.al.\/} with $\alpha(t) = \alpha_0 t$, $\beta(t) = \beta_0 \sin(\omega
t)$ and $\chi=\pi/2$ \cite{Shi65}.

In conclusion, we have shown that the fields (\ref{1.3}), derived by using
inverse techniques in \cite{Fer97}, produce exactly solvable spin-1/2
Hamiltonians which, at first sight, should be solved by approximate or
numerical methods. We have shown also that the time evolution induced by
these Hamiltonians can produce evolution loops, which simplify the
calculation of the integrals involved in the formula for the corresponding
geometric phases. 

This work has been supported by CONACYT, Mexico (ref 973035 and project
26329-E) and Junta de Castilla y Le\'on, Spain (project C02/97). ORO wants
to thank to L.M.~Nieto for his patience during the long discussions about
this paper, and to the colleagues at Departamento de F\'{\i}sica Te\'orica
by their interest in this work.



\end{document}